\documentstyle[prl,aps,preprint]{revtex}

\begin{document}

\widetext
\draft
\title{
The Kosterlitz-Thouless Transitions on Fluctuating Surface of Genus Zero
}
\author{Jeong-Man Park}
\address{
Department of Physics, 
University of Pennsylvania, Philadelphia, PA 19104
}
\maketitle

\begin{abstract}
We investigate the Kosterlitz-Thouless transition for hexatic order on
a fluctuating spherical surface of genus zero and derive a Coulomb
gas Hamiltonian to describe it.
In the Coulomb gas Hamiltonian, charge densities arises from
disclinations and from Gaussian curvature.
There is an interaction coupling the difference between these two densities,
whose strength is determined by the hexatic rigidity.
We then convert it into the sine-Gordon Hamiltonian and find a linear coupling
between a scalar field and the Gaussian curvature.
After integrating over the shape fluctuations, we obtain the massive
sine-Gordon Hamiltonian, which corresponds to a neutral Yukawa gas,
and the interaction between the disclinations is screened.
We find, for $K_{A}/\kappa \gg 1/2$ where $K_{A}$ and $\kappa$ are
hexatic and bending rigidity, respectively, the transition is
supressed altogether, much as the Kosterlitz-Thouless transition is
supressed in an infinite 2D superconductor.
If on the other hand $K_{A}/\kappa \ll 1/2$, there can be an effective
transition.
\end{abstract}
\clearpage
\par
Recently, the Kosterlitz-Thouless (KT) transition for hexatic order on a
free fluctuating membrane has been investigated \cite{jp-lub}.
A flat rigid membrane can have quasi-long-range (QLR) hexatic order 
\cite{Nelson79}
at low temperature and undergo a KT disclination unbinding transition 
\cite{Kosterlitz72,Nelson83,Minnhagen87}
to a disordered high-temperature phase.
A fluctuating membrane can also have QLR hexatic order 
\cite{Nelson87} at low temperature.
At high temperature, a fluctuating membrane has no internal order and
can be charaterized by a bending rigidity $\kappa$.
At length scales less than the persistence length 
$\xi_{p}= ae^{4\pi\kappa/3T}$, where $a$ is a molecular size and $T$ is
the temperature, the membrane looks flat; at longer length scales, it is
crumpled.
However, at low temperature, hexatic order stiffens the bending rigidity
so that the bending rigidity approaches a constant times the hexatic rigidity
$K_{A}$ \cite{David87}.
Thus, the hexatic membrane is more rigid than a fluid membrane, and it is said 
to be crinkled rather than crumpled.
A fluctuating hexatic membrane can undergo a KT transition from the crinkled
hexatic to the crumpled fluid state.
For fixed large $\kappa$, there is a disclination melting to the crumpled
fluid phase as temperature is increased, and at fixed $K_{A}$, there is a
transition to the crumpled fluid phase as $\kappa$ is decreased.
\par
In this paper, we extend a study of the KT transitions to a fluctuating
surface of genus zero.
Ovrut and Thomas discussed the structure of the KT transition of a 
vortex-monopole Coulomb gas on a rigid sphere and show it is the same as 
in the planar case, {\it i.e.} the KT transition temperature on a rigid
sphere is the same as that on the Euclidean plane ; $T^{\rm KT}_{\rm sphere}
= T^{\rm KT}_{\rm plane} = \pi/2K_{A}$ \cite{Ovrut}.
We investigate the effect of thermal shape fluctuations of a genus zero
surface on the KT transition in the limit $\beta\kappa \gg 1$.
In this limit, we can parametrize the surface by its radius vector
as a function of standard polar coordinates ${\bf u}=(\theta,\phi)
\equiv \Omega$;
\begin{equation}
{\bf R}(\Omega) = R(1+\rho(\Omega)){\bf e}_{r},
\end{equation}
where ${\bf e}_{r}$ is the radial unit vector and $\rho(\Omega)$ measures
deviation from sphericity with radius $R$.
This parametrization is a ``normal gauge''.
To make the equations simple, we map this parametrization onto the
unit sphere parametrization with $R=1$.
Later when we analyze interaction between two disclinations,
we will recover this length scale.
Associated with ${\bf R}(\Omega)$ is a metric tensor $g_{\alpha\beta}(\Omega)=
\partial_{\alpha}{\bf R}(\Omega)\cdot\partial_{\beta}{\bf R}(\Omega)$ 
and a curvature tensor $K_{\alpha\beta}(\Omega)$ defined via 
$K_{\alpha\beta}(\Omega)= {\bf N}(\Omega)\cdot\partial_{\alpha}
\partial_{\beta}{\bf R}(\Omega)$, 
where ${\bf N}(\Omega)$ is the local unit normal to the surface.
From the curvature tensor $K_{\alpha\beta}$, the mean curvature, $H$, and the 
Gaussian curvature, $K$, are defined as follows:
\begin{equation}
H = \frac{1}{2} g^{\alpha\beta}K_{\beta\alpha}, \;\; 
K = \det g^{\alpha\lambda}K_{\lambda\beta},
\end{equation}
where $g^{\alpha\beta}$ is the inverse tensor of $g_{\alpha\beta}$ 
satisfying $g^{\alpha\lambda}g_{\lambda\beta}=\delta^{\alpha}_{\beta}$.
\par
To describe hexatic order, we construct the tangent vectors
\begin{equation}
{\bf t}_{\theta} = \partial_{\theta}{\bf R},\;\; 
{\bf t}_{\phi} = \partial_{\phi}{\bf R},
\end{equation}
where $\partial_{\alpha}=\partial/\partial u^{\alpha}$, ${\bf u}=
(\theta,\phi)$,
and introduce orthonormal unit vectors ${\bf e}_{1}$ and ${\bf e}_{2}$
at each point on the surface.
Then, ${\bf e}_{1}(\Omega)\cdot {\bf t}_{\theta}(\Omega) = \cos\Theta(\Omega)$ 
defines a local bond angle $\Theta(\Omega)$.
Hexatic order is then described by the local bond order parameter 
${\bf m}(\Omega) = \cos\Theta {\bf e}_{1} + \sin\Theta {\bf e}_{2}$, 
where $\Theta(\Omega)$ has 6-fold symmetry.
Note that since $\Theta(\Omega)$ depends on the choice of orthonormal vectors
${\bf e}_{1}$ and ${\bf e}_{2}$, any spatial derivatives for ${\bf m}$ must be 
covariant derivatives.
\par
In the continuum elastic theory, it is now well established that the 
long-wavelength properties of a fluctuating hexatic membrane are described
by the Helfrich-Canham Hamiltonian ${\cal H}_{\rm HC}$ \cite{Helf-Canh}
and the hexatic free energy ${\cal H}_{\rm A}$ \cite{Nelson87}.
The Helfrich-Canham Hamiltonian can be expressed as a sum of three terms,
\begin{equation}
{\cal H}_{\rm HC} = {\cal H}_{\kappa} + {\cal H}_{\rm G} + {\cal H}_{\sigma}.
\end{equation}
The first term is the mean curvature energy,
\begin{eqnarray}
{\cal H}_{\kappa} & = & \frac{1}{2}\kappa\int d^{2}u\sqrt{g} (2H-2H_{0})^{2}
                        \nonumber  \\
                  & = & \frac{1}{2}\kappa\int d\Omega
\left( (\nabla^{2} + 2)\rho \right)^{2} + {\cal O}(\rho^{3})
\label{non-lin-cur}
\end{eqnarray}
where $\sqrt{g} = \det g_{\alpha\beta}$,
$2H=K^{\alpha}_{\alpha}$ is the twice the mean curvature, $H_{0}$ is
the spontaneous mean curvature which is equal to 1 for the sphere,
and the second 
form is valid for the normal gauge.
The second term is the Gaussian curvature energy,
\begin{equation}
{\cal H}_{\rm G} = \frac{1}{2}\kappa_{G}\int d^{2}u\sqrt{g} K,
\end{equation}
where $K = \det K^{\alpha}_{\beta}$ is the Gaussian curvature. 
This term is a topological 
invariant depending only on the genus of the surfaces due to the Gauss-Bonnet
theorem;
\begin{equation}
\int d^{2}u\sqrt{g} K = 4\pi(1-\eta) = 2\pi\chi,
\label{gauss-bon}
\end{equation}
where $\eta$ is the number of handles and $\chi=2(1-\eta)$ is the
Euler characteristic.
Since we will consider surfaces of fixed genus, we will drop this term.
Finally
\begin{equation}
{\cal H}_{\sigma} = \sigma\int d^{2}u\sqrt{g}
\end{equation}
is the surface tension energy. We are mostly interested in free membranes
for which the renormalized surface tension obtained by differentiating
the total free energy ${\cal F}$ with respect to the total surface area
${\cal A}$ $(\sigma_{R} = \partial {\cal F}/\partial {\cal A})$, is zero.
Since there are entropic contributions to $\sigma_{R}$ as well as contributions
from internal order, the value of the bare surface tension $\sigma$ will
have to be adjusted to keep $\sigma_{R}$ zero.
In what follows, we will ignore ${\cal H}_{\sigma}$ with the understanding
that it is really present if we want to keep track of how $\sigma_{R}$
actually becomes zero.
\par
The hexatic free energy is the contribution to the energy from
fluctuations in the local bond order parameter.
Since the hexatic order parameter ${\bf m}$ has a fixed magnitude and there 
are 
no external fields aligning ${\bf m}$ along a particular direction, the lowest
nontrivial contribution to the energy associated with ${\bf m}$ arises from
its gradients,
\begin{equation}
{\cal H}_{A} = \frac{1}{2}K_{A} \int d^{2}u \sqrt{g}
g^{\alpha\beta} D_{\alpha}{\bf m} \cdot D_{\beta}{\bf m},
\end{equation}
where $D_{\alpha}$ is a covariant derivative since 
the bond order parameter is frustrated by the rotation of tangent vectors
that occurs under parallel transport on a curved surface. The amount of
frustration is given by the gauge field $A_{\alpha}$, {\it i.e.} the covariant
derivative of ${\bf e}_{a}$ in direction $\alpha$ defines the gauge field
$A_{\alpha}$.
Under parallel transport in direction $du^{\alpha}$, each ${\bf e}_{a}$ is 
rotated by an angle $A_{\alpha} du^{\alpha}$.
Thus the gauge field $A_{\alpha}$ is defined by
\begin{equation}
D_{\alpha} {\bf e}_{a} = -A_{\alpha}\varepsilon_{ab}{\bf e}_{b},
\end{equation}
where $\varepsilon_{ab}$ is the antisymmetric tensor with
$\varepsilon_{12}=-\varepsilon_{21}=1$
and $A_{\alpha}\varepsilon_{ab}$ is called the spin-connection and describes
how the basis vector ${\bf e}_{a}$ rotates under parallel transport according
to the Gaussian curvature $K$ of the surface.
In fact, $A_{\alpha}$ is related to $K$.
The curl of the gauge field $A_{\alpha}$ is the Gaussian curvature;
\begin{equation}
\gamma^{\alpha\beta}D_{\alpha}A_{\beta} = K,
\label{curl-spin}
\end{equation}
where $\gamma^{\alpha\beta}$ is the antisymmetric tensor defined via
\begin{eqnarray}
\gamma_{\alpha\beta} & = & {\bf N}\cdot ({\bf t}_{\alpha}
\times{\bf t}_{\beta}) =
\sqrt{g}\varepsilon_{\alpha\beta},  \nonumber  \\
\gamma^{\alpha\beta} & = & g^{\alpha\alpha'}g^{\beta\beta'}
\gamma_{\alpha'\beta'},
\label{antisym}
\end{eqnarray}
\par
In terms of the local bond angle $\Theta(\Omega)$, 
the covariant derivative of ${\bf m}$ writes
\begin{eqnarray}
D_{\alpha} {\bf m} & = & (D_{\alpha} m_{a}) {\bf e}_{a} +
                         m_{a} (D_{\alpha}{\bf e}_{a})  \nonumber  \\
   & = & (D_{\alpha} m_{a}) {\bf e}_{a} -
         m_{a} A_{\alpha}\varepsilon_{ab} {\bf e}_{b}  \nonumber  \\
   & = & (D_{\alpha}\Theta) (-\sin\Theta {\bf e}_{1} + \cos\Theta {\bf e}_{2})
  \nonumber  \\
   &   & - A_{\alpha} (\cos\Theta {\bf e}_{2} - \sin\Theta {\bf e}_{1}) 
  \nonumber \\
   & = & (D_{\alpha}\Theta - A_{\alpha}) {\bf m}_{\perp},
\end{eqnarray}
where ${\bf m}_{\perp} = -\sin\Theta {\bf e}_{1} + \cos\Theta {\bf e}_{2}$
satisfying ${\bf m}\cdot{\bf m}_{\perp} = 0$.
Then the hexatic energy writes
\begin{equation}
{\cal H}_{A}  = \frac{1}{2}K_{A} \int d^{2}u \sqrt{g}
g^{\alpha\beta}(\partial_{\alpha}\Theta-A_{\alpha}) 
(\partial_{\beta}\Theta-A_{\beta}).
\end{equation}
Thus we have the Hamiltonian ${\cal H}={\cal H}_{\kappa}+{\cal H}_{\rm A}$
to describe fluctuating hexatic membranes.
\par
To gain some physical understanding of a spherical hexatic membrane, 
we examine the ground states.
Since we are interested in the limit $\beta\kappa \gg 1$ and
in this limit ${\cal H}_{\kappa}$ dominates, we first minimize
${\cal H}_{\kappa}$ over the shape fluctuation field $\rho$ which gives
$\rho(\Omega) = 0$ and
then we minimize ${\cal H}_{A}$ over $\Theta$ with $\rho(\Omega)=0$ and find
\begin{equation}
\left. \frac{\delta{\cal H}^{0}_{A}}{\delta\Theta(\Omega)} 
\right|_{\Theta=\Theta^{0}} = \frac{1}{\sqrt{g^{0}}} \partial_{b}
{g^{0}}^{ab} (\partial_{a}\Theta^{0}-A^{0}_{a}) = 0,
\end{equation}
where the superscript$~^{0}$ stands for the rigid sphere with 
$\rho(\Omega)=0$.
In Ref.~\cite{Lubensky92}, Lubensky and Prost show that in the ground state
12 disclinations of strength $2\pi/6$ are arranged at the vertices of 
icosahedron inscribed in the sphere.
A disclination at ${\bf u} = {\bf u}_{i}$ with 
strength $q_{i}$ gives rise to a singular contribution $\Theta^{\rm sing}_{0}$
to $\Theta^{0}$ satisfying 
\begin{equation}
\oint_{\Gamma} du^{\alpha}\partial_{\alpha}\Theta^{\rm sing}_{0} = q_{i},
\end{equation}
where $\Gamma$ is a contour enclosing ${\bf u}_{i}$.
Thus, in general $\partial_{\alpha}\Theta^{0} = \partial_{\alpha}\Theta'_{0}
+v^{0}_{\alpha}$ where $\Theta'_{0}$ is nonsingular, $v^{0}_{\alpha}=
\partial_{\alpha}\Theta^{\rm sing}_{0}$, and
\begin{equation}
\gamma^{\alpha\beta}D_{\alpha}v^{0}_{\beta} = n^{0}(\Omega),
\end{equation}
where
\begin{equation}
n^{0}(\Omega) = \frac{2\pi}{6} \sum^{12}_{i=1} 
\delta(\Omega-\Omega_{i}),
\end{equation}
where is the disclination density in the ground state and
$\Omega_{i}$'s are the coordinates of the vertices of icosahedron.
Since $(\partial_{\alpha}\Theta^{0}-A_{\alpha}^{0})$ satisfies 
$D^{\alpha}(\partial_{\alpha}\Theta^{0}-A_{\alpha}^{0}) = 0$,
it is divergence-less and purely transverse.
Accordingly $(\partial_{\alpha}\Theta^{0}-A_{\alpha}^{0})$ can be written
in terms of the curl of scalar fields and by applying the operator 
$\gamma^{\beta\alpha} D_{\beta}$ to 
$(\partial_{\alpha}\Theta^{0}-A^{0}_{\alpha})$, we find
these scalar fields to be related to the Gaussian curvature $K_{0}$ of
the rigid sphere and the ground state disclination density on the rigid
sphere,
\begin{eqnarray}
\gamma^{\beta\alpha} D_{\beta}
(\partial_{\alpha}\Theta^{0}-A^{0}_{\alpha}) & = & 
    \gamma^{\beta\alpha}D_{\beta}v^{0}_{\alpha} - 
    \gamma^{\beta\alpha}D_{\beta}A^{0}_{\alpha}  \nonumber  \\
   & = & n^{0} - K_{0},
\end{eqnarray}
where $K_{0}$ is a Gaussian curvature of the rigid sphere and
$n^{0}$ is the disclination density in the ground state.
\par
Now taking into account the bond angle fluctuations around $\Theta^{0}$ and
the shape fluctuations around the sphere, 
\begin{equation}
\Theta = \Theta^{0} + \tilde{\Theta}, \;\;
A_{\alpha} = A^{0}_{\alpha} + \delta A_{\alpha},
\end{equation}
the full
Hamiltonian writes ${\cal H}={\cal H}_{0}+\delta{\cal H}$
\begin{eqnarray}
{\cal H}_{0} & = & \frac{1}{2}K_{A} \int d\Omega 
(\partial^{\alpha}\Theta^{0}-{A^{0}}^{\alpha})(\partial_{\alpha}\Theta^{0}
-A^{0}_{\alpha})
\nonumber  \\
\delta{\cal H} & = & \frac{1}{2}\kappa \int d\Omega \left( (\nabla^{2}+2)\rho
\right)^{2}
\nonumber  \\
   &   & + \frac{1}{2}K_{A} \int d\Omega
(\partial^{\alpha}\tilde{\Theta}-\delta A^{\alpha})
(\partial_{\alpha}\tilde{\Theta}-\delta A_{\alpha}) + {\cal O}(\rho^{3}).
\end{eqnarray}
The angle fluctuation field $\tilde{\Theta}(\Omega)$ can also have 
disclinations of strength $q=2\pi(k/6)$ where $k$ is an integer,
due to the thermal fluctuation \cite{Lubensky95}.
Thus, $\partial_{\alpha}\tilde{\Theta}$ can be decomposed into singular
and nonsingular parts;
$\partial_{\alpha}\tilde{\Theta} = \partial_{\alpha}\Theta^{\parallel} + 
v_{\alpha}$ where $\Theta^{\parallel}$ is
nonsingular, $v_{\alpha} = \partial_{\alpha}\tilde{\Theta}^{\rm sing}$ and
\begin{equation}
\gamma^{\alpha\beta}D_{\alpha} v_{\beta} = n(\Omega),\;\;
n(\Omega) = \sum_{i} q_{i} \delta(\Omega - \Omega_{i}),
\label{eq:vortex}
\end{equation}
where $n(\Omega)$ is the thermally-excited disclination density
with disclinations of strength $q_{i}$ at $\Omega_{i}$.
The vector $v_{\alpha}$ can always be chosen so that it is purely transverse,
{\it i.e.} $D_{\alpha}v^{\alpha} = 0$. 
In the hexatic Hamiltonian, $\partial_{\alpha}\tilde{\Theta}$ always occurs 
in the combination
$(\partial_{\alpha}\tilde{\Theta} - \delta A_{\alpha})$. 
The spin-connection $\delta A_{\alpha}$ can and will
in general have both a longitudinal and a transverse component.
However, one can always redefine $\Theta^{\parallel}$ to include the 
longitudinal part of $\delta A_{\alpha}$. This amounts to choosing 
locally rotated orthonormal vectors
${\bf e}_{1}({\bf u})$ and ${\bf e}_{2}({\bf u})$ so that 
$D_{\alpha} \delta A^{\alpha} = 0$.
Thus we may take both $v_{\alpha}$ and $\delta A_{\alpha}$ to be transverse and
the hexatic Hamiltonian
\begin{eqnarray}
   &   & \frac{1}{2}K_{A} \int d\Omega 
(\partial^{\alpha} \Theta^{\parallel}+v^{\alpha} -\delta A^{\alpha})
(\partial_{\alpha} \Theta^{\parallel}+v_{\alpha} -\delta A_{\alpha})
\nonumber  \\
   & = & {\cal H}_{\parallel} + {\cal H}_{\perp},
\end{eqnarray}
can be decomposed into a regular longitudinal part,
\begin{equation}
{\cal H}_{\parallel} =  \frac{1}{2}K_{A}\int d\Omega
\partial^{\alpha}\Theta^{\parallel}\partial_{\alpha}\Theta^{\parallel},
\end{equation}
and a transverse part,
\begin{equation}
{\cal H}_{\perp} =  \frac{1}{2}K_{A}\int d\Omega
(v^{\alpha} - \delta A^{\alpha})(v_{\alpha} - \delta A_{\alpha}),
\label{eq:h-perp}
\end{equation}
where the cross term $\int d\Omega 
(v_{\alpha}-\delta A_{\alpha}) \partial^{\alpha}\Theta^{\parallel}$ 
is dropped since
$D^{\alpha}(v_{\alpha}-A_{\alpha}) = 0$.
\par
It costs an energy $\epsilon_{c}(k)$ to create the core of a disclination
of strength $k$. (We assume for the moment that the core energies of the
positive and negative disclinations are the same.  See, however, Refs.
\cite{jp-lub,jp-lub3} and Ref.\cite{Nelson96}.) 
Thus, partition sums should be weighted by a factor
$y_{k} = e^{-\beta\epsilon_{c}(k)}$ for each disclination of strength
$k$. Since $\epsilon_{c}(k) \sim k^{2}$, we may at low temperature
restrict our attention to configurations in which only configurations
of strength $\pm 1$ appear. Let $N_{\pm}$ be the number of disclinations
of strength $\pm 1$ and let ${\bf u}_{i^{\pm}}$ be the coordinate of the 
core of
the disclination with strength $\pm 1$ labeled by $i$. The hexatic membrane
partition function can then be written as
\begin{equation}
{\cal Z} (\kappa, K_{A}, y) = {\rm Tr}_{\rm v} y^{N}
\int{\cal D}{\bf R}\int{\cal D}\Theta^{\parallel}
e^{-\beta\delta{\cal H}_{\kappa}}e^{-\beta({\cal H}_{\parallel}+
{\cal H}_{\perp})},
\label{eq:partition}
\end{equation}
where $y=y_{1}$, and $N = N_{+} + N_{-}$.
${\cal H}_{\perp}$ depends on all of the disclination coordinates
$\Omega_{\nu^{\pm}}$ where $\nu^{\pm}=1,2,\cdots,N_{\pm},$ and
${\rm Tr}_{\rm v}$ is the sum over all possible disclination
distribution with the topological 
constraint \cite{Spivak79};
\begin{equation}
{\rm Tr}_{\rm v} = \sum_{N_{+},N_{-}} \frac{\delta_{N_{+},N_{-}}}
{N_{+}!N_{-}!} \prod_{\nu^{+}}\int \frac{d\Omega_{\nu^{+}}}{a^{2}}
\prod_{\nu^{-}}\int \frac{d\Omega_{\nu^{-}}}{a^{2}},
\end{equation}
where $a^{2}$ is a molecular solid angle. 
The Kronecker factor $\delta_{N_{+},N_{-}}$
in ${\rm Tr}_{\rm v}$ imposes the topological constraint that the total 
disclination strength on a sphere is 2 since with $N_{+}=N_{-}$ we have 12 
ground state disclinations with the strength 1/6 giving the total disclination
strength $12 \times (1/6) =2$.
\par
The hexatic model of Eq.~(\ref{eq:partition}) can easily be converted to a 
Coulomb gas model using
\begin{equation}
\gamma^{\alpha\beta}D_{\alpha}(v_{\beta}-\delta A_{\beta}) = 
n - \delta K \equiv {\cal C},
\label{eq:curl-vel}
\end{equation}
which follows from Eq.~(\ref{curl-spin}) and Eq.~(\ref{eq:vortex}) where
$\delta K$ is the deviation of the Gaussian curvature from the rigid sphere.
The quantity ${\cal C} = n-\delta K$ is a ``charge'' density 
with contributions arising 
both from disclinations and from Gaussian curvature.
Equation (\ref{eq:curl-vel}) implies
\begin{equation}
v_{\alpha}-\delta A_{\alpha} = -{\gamma_{\alpha}}^{\beta}D_{\beta} 
\frac{1}{\Delta}{\cal C},
\label{eq:va-Aa}
\end{equation}
where we used $\gamma_{\alpha\lambda}D^{\lambda}\gamma^{\alpha\beta}D_{\alpha}
= - \Delta$ and 
$\Delta=D^{\alpha}D_{\alpha} = (1/\sqrt{g})\partial_{\alpha}\sqrt{g}
g^{\alpha\beta}\partial_{\beta}$ 
is the Laplacian on a surface with metric tensor 
$g_{\alpha\beta}$ acting on a scalar. 
Recall [Eq.~(\ref{antisym})] that ${\gamma_{\alpha}}^{\beta}$ rotates
a vector by $\pi/2$ so that $(v_{\alpha}-\delta A_{\alpha})$ is perpendicular 
to
$D_{\beta}(-\Delta)^{-1}{\cal C}$ and is thus manifestly transverse.
Using Eq.~(\ref{eq:va-Aa}) in Eq.~(\ref{eq:h-perp}), we obtain
\begin{equation}
{\cal Z} = {\rm Tr}_{\rm v} y^{N}\int{\cal D}{\bf R}
\int{\cal D}\Theta^{\parallel} e^{-\beta
{\cal H}_{\kappa} -\beta{\cal H}_{\parallel} -\beta{\cal H}_{\rm c}},
\end{equation}
where
\begin{eqnarray}
{\cal H}_{\rm c} & = & \frac{1}{2}K_{A} \int d\Omega 
\frac{\gamma_{\alpha}^{\beta}D_{\beta}}{\Delta} {\cal C}(\Omega)
\frac{\gamma^{\alpha\lambda}D_{\lambda}}{\Delta}{\cal C}(\Omega)  \nonumber  \\
   & = & \frac{1}{2}K_{A} \int d\Omega d\Omega'
{\cal C}(\Omega) \left( \frac{\gamma_{\alpha}^{\beta}D_{\beta}
\gamma^{\alpha\lambda}D_{\lambda}}{\Delta^{2}} \delta(\Omega-\Omega')
\right) {\cal C}(\Omega')   \nonumber  \\
   & = & \frac{1}{2}K_{A} \int d\Omega d\Omega'
{\cal C}(\Omega) \left( -\frac{1}{\Delta} \delta(\Omega-\Omega') \right)
{\cal C}(\Omega'),
\label{coulomb-rr}
\end{eqnarray}
is the Coulomb Hamiltonian associated with the charge ${\cal C}$.
Since the longitudinal variable $\Theta^{\parallel}$ appears only 
quadratically in
${\cal H}_{\parallel}$, the trace over $\Theta^{\parallel}$ can be done 
directly giving the Liouville action \cite{Polyakov81} 
arising from the conformal anomaly;
\begin{equation}
\int {\cal D}\Theta^{\parallel} e^{-\beta{\cal H}_{\parallel}} = 
e^{-\beta{\cal H}_{\rm L}},
\end{equation}
where
\begin{equation}
\beta{\cal H}_{\rm L} = \frac{1}{8\pi a^{2}}\int d\Omega -
\frac{1}{24\pi}\int d\Omega d\Omega' K(\Omega)
\left(-\frac{1}{\Delta} \delta(\Omega-\Omega')\right) K(\Omega').
\end{equation}
\par
The Coulomb gas partition function can thus be written
\begin{equation}
{\cal Z} = {\rm Tr}_{\rm v} y^{N}\int{\cal D}{\bf R} 
e^{-\beta{\cal H}_{\kappa}
-\beta{\cal H}_{\rm L} -\beta{\cal H}_{\rm C}}.
\label{eq:tr-hs}
\end{equation}
The Coulomb gas model can be converted following standard procedures
into a sine-Gordon model. The first step is to carry out a 
Hubbard-Stratonovich transformation on $\beta{\cal H}_{\rm C}$:
\begin{equation}
e^{-\beta{\cal H}_{\rm C}} = e^{\beta{\cal H}_{\rm L}}
\int{\cal D}\Phi e^{-\frac{1}{2}(\beta K_{A})^{-1}
\int d\Omega \partial^{\alpha}\Phi
\partial_{\alpha}\Phi} e^{i\int d\Omega {\cal C}\Phi},
\end{equation}
where the Liouville factor $e^{\beta{\cal H}_{\rm L}}$ is needed to ensure
that $e^{-\beta{\cal H}_{\rm C}}$ be one when ${\cal C}=0$. 
Inserting this in Eq.~(\ref{eq:tr-hs}), we obtain
\begin{equation}
{\cal Z} = {\rm Tr}_{\rm v} y^{N}\int{\cal D}{\bf R}{\cal D}\Phi 
e^{-\beta{\cal H}_{\kappa}-\beta{\cal H}_{\Phi}}
e^{i\int d\Omega (n-\delta K)\Phi},
\end{equation}
where 
\begin{equation}
\beta{\cal H}_{\Phi} = \frac{1}{2}(\beta K_{A})^{-1}\int d\Omega
\partial^{\alpha}\Phi\partial_{\alpha}\Phi.
\end{equation}
The only dependence on disclinations is now in the term linear in $n$. 
Thus to carry
out ${\rm Tr}_{\rm v}$, we need only to evaluate
\begin{eqnarray}
   &   & {\rm Tr}_{\rm v} y^{N}e^{i\int d\Omega n\Phi}   \nonumber   \\
   & = & \sum_{N_{+},N_{-}}\frac{1}{N_{+}!N_{-}!}\delta_{N_{+},N_{-}}
y^{N_{+}+N_{-}} \left( \int\frac{d\Omega}{a^{2}} 
e^{2\pi i\Phi(\Omega)/6}\right)^{N_{+}}  
\left( \int\frac{d\Omega}{a^{2}} 
e^{-2\pi i\Phi(\Omega)/6}\right)^{N_{-}}   \nonumber  \\
   & = & \sum_{N_{+},N_{-}}\frac{1}{N_{+}!N_{-}!}
\int \frac{d\omega}{2\pi} 
\left( y \int\frac{d\Omega}{a^{2}} 
e^{i \{ 2\pi [\Phi(\Omega)/6] -\omega \} }\right)^{N_{+}}    
\left( y \int\frac{d\Omega}{a^{2}} 
e^{-i \{ 2\pi [\Phi(\Omega)/6] -\omega \} }\right)^{N_{-}}     \nonumber  \\
   & = & \int \frac{d\omega}{2\pi} 
e^{(2y/a^{2})\int d\Omega \cos[2\pi(\Phi/6)-\omega]}.
\end{eqnarray}
Thus
\begin{equation}
{\cal Z} = \int \frac{d\omega}{2\pi}\int{\cal D}\Phi\int{\cal D}{\bf R}
               e^{-\beta{\cal H}_{\kappa}}e^{-\beta{\cal H}_{\Phi}}
               e^{(2y/a^{2})\int d\Omega
               \cos[2\pi(\Phi/6)-\omega]} e^{-i\int d\Omega
               \Phi \delta K}.
\label{eq:tr-sg}
\end{equation}
We can now change variables, letting $\Phi = (6/2\pi)(\Phi'+\omega)$.
The term linear in the Gaussian curvature then becomes
\begin{equation}
-i\int d\Omega \delta K \frac{6}{2\pi}(\omega + \Phi')
= -i\frac{p}{2\pi}\int d\Omega \Phi' \delta K,
\end{equation}
where we used $\int d\Omega \delta K = 0$.
The integral over $\omega$
in Eq.~(\ref{eq:tr-sg}) is now trivial, and dropping the prime we obtain
\begin{equation}
{\cal Z} = \int{\cal D}\Phi\int{\cal D}{\bf R} e^{-\beta{\cal H}_{\kappa}}
e^{-{\cal L}},
\label{eq:par-sg}
\end{equation}
where 
\begin{eqnarray}
{\cal L} & = & \frac{1}{2}(\beta K_{A})^{-1}\left(\frac{6}{2\pi}\right)^{2}
\int d\Omega \partial^{\alpha}\Phi\partial_{\alpha}\Phi 
\nonumber  \\  
        &   & -\frac{2y}{a^{2}}\int d\Omega \cos\Phi 
        -i\frac{6}{2\pi}\int d\Omega \Phi \delta K
\end{eqnarray}
is the sine-Gordon action on a fluctuating surface of genus zero.
The first two terms of this action are the gradient and cosine energies
present on a rigid sphere. The final term provides the principal coupling
between $\Phi$ and fluctuations in the metric.
It is analogous to the dilaton coupling \cite{Green87} of string theory 
though here
the coupling constant is imaginary rather than real.
Note that the Liouville action is not explicitly present in 
Eq.~(\ref{eq:par-sg}). 
\par
In the regime $\beta\kappa \gg 1$, we can truncate the higher order terms
in $\rho$. In the normal gauge, the partition function becomes
\begin{eqnarray}
{\cal Z} & = & \int {\cal D}\rho{\cal D}\Phi
   \exp\left[-\frac{1}{2}\beta\kappa\int d\Omega
               \left((\nabla^{2}+2)\rho\right)^{2}  \right. \nonumber  \\
         &   & -\frac{1}{2}\beta\Gamma\int d\Omega (\nabla\Phi)^{2} +
   \frac{2y}{a^{2}}\int d\Omega\cos\Phi                   \nonumber  \\
         &   & \left. +i\frac{6}{2\pi}\int d\Omega \Phi(\nabla^{2}+2)\rho 
   \right]
\end{eqnarray}
where 
$\beta\Gamma \equiv 36/(4\pi^{2}\beta K_{A})$ and we used 
$\delta K = (\nabla^{2}+2)\rho$.      
To lowest order in $\rho$, the shape fluctuation field $\rho$ is linearly
coupled to the scalar field $\Phi$ which is the conjugate field to the
disclinations.
In Ref.~\cite{jp-lub}, we have shown the similar coupling in the
fluctuating flat membrane is quadratic in the shape fluctuation field.
\par
Integrating over the shape fluctuation field $\rho$ gives
the effective Hamiltonian for the conjugate field to the disclinations
\begin{eqnarray}
{\cal Z} & = & \int {\cal D}\Phi\exp\left[-\frac{1}{2}\beta\Gamma\int d\Omega
       \left((\nabla\Phi)^{2} + \mu^{2}\Phi^{2}\right) \right.  \nonumber  \\
         &   &  \left. + \frac{2y}{a^{2}}\int d\Omega\cos\Phi \right]
\end{eqnarray}
with $\mu^{2}=K_{A}/\kappa$.
This is the massive sine-Gordon theory. The shape 
fluctuations induce the mass term for $\Phi$ field and screen the Coulombic
interaction between the disclinations giving the Yukawa interaction between
them.
This partition function is equivalent to that of the Yukawa gas Hamiltonian
on the rigid sphere with radius $R$;
\begin{eqnarray}
{\cal H}_{\rm Yukawa} & = & \frac{1}{2} \beta K_{A} \int d\Omega
n(\Omega) \frac{1}{(-\nabla^{2}+\mu^{2})} n(\Omega) \nonumber  \\
   & = & \frac{1}{2} \beta K_{A} \sum_{i,j} q_{i}q_{j}
G(\Omega_{i}-\Omega_{j}),
\end{eqnarray}
where
\begin{eqnarray}
G(\Omega_{i}-\Omega_{j}) & = & \sum_{l} 
\frac{2l+1}{l(l+1)+\mu^{2}}P_{l}(\cos\omega_{ij}) \nonumber  \\
   & = & -\frac{\pi}{\cos\left( (\nu+\frac{1}{2})\pi \right)}
P_{\nu}(-\cos\omega_{ij}),
\end{eqnarray}
where $P_{\nu}(\cdot)$ is the Legendre polynomial with degree 
$\nu=-1/2 \pm (\sqrt{1-4\mu^{2}})/2$ and
$\omega_{ij}$ is the angle between two disclinations at $\Omega_{i}$ and
$\Omega_{j}$.
For $0 \leq \mu \leq 1/2$, degree of Legendre polynomial, $\nu$, is real
and the length scale introduced by $\lambda_{d} \equiv (\mu/R)^{-1}$ is larger
than the system size, $2R$ after recovering the original length scale
by mapping the unit sphere back to the sphere with the radius $R$.
On the other hand, if $\mu > 1/2$, $\nu$ is a complex number,
$\nu=-1/2 \pm i\tau$ where $\tau = (\sqrt{4\mu^{2}-1})/2$ and
$\lambda_{d} < 2R$.
The length scale introduced by $\lambda_{d}=R/\mu$ may be interpreted 
as the Debye screening length arising from shape fluctuations.
\par
The interaction energy between two disclinations $i$ and $j$ at positions
$\Omega_{i}$ and $\Omega_{j}$ with strength $q_{i}$ and $q_{j}$ is given by
$q_{i}q_{j}G(d_{ij})$ where $d_{ij}=2R\sin(\omega_{ij}/2)$ is 
the chordal distance between two disclinations on the sphere with radius $R$.
The interaction $G(d_{ij})$ has the following limiting forms:
\begin{equation}
G(d_{ij}) \simeq \left\{ \begin{array}{ll}
        -\frac{1}{2} \ln \left( d_{ij}/2 \right), &
                  \mu^{-1} \gg 2,\; d_{ij} \ll 2R  \\
        -\frac{1}{2} \ln \left( d_{ij}/2 \right), &
                  \mu^{-1} \ll 2,\; d_{ij} \ll \lambda_{d}  \\
        e^{-d_{ij}/\lambda_{d}},  &
                  \mu^{-1} \ll 2,\; d_{ij} \gg \lambda_{d}
                         \end{array}
                 \right.
\end{equation}
Following the analogy of the 2D Coulomb gas, when the screening length is 
much larger than the system size, $\lambda_{d} \gg 2R$, the induced mass term 
arising from shape fluctuations is irrelavent for the KT transition
and the KT transition temperature is given by
$T_{\rm c} = \frac{\pi}{72K_{A}}$ for $\mu \ll 1/2$.
However, for $\mu \gg 1/2$, the screening length is 
shorter than the system size,
$\lambda_{d} \ll 2R$, and the mass term is relavent for the KT transition and
changes the universality class of the system.
There is no KT transition at non-zero temperature.
The disclinations are always unbound at non-zero temperature and
the KT transition temperature vanishes.
Thus we find the crossover at $\mu = 1/2$.
\par
In conclusion, we present the effect of shape fluctuations
on the interaction of the disclinations on a spherical surface with
genus zero.
We have confirmed that the screened interaction is of the same form as 
the vortex line interactions in type-II superconductors. 
In these superconductors, screening of vortex line interactions drives the 
Kosterlitz-Thouless transition temperature to zero for an infinite 
superconductor in zero magnetic field \cite{Moore89}.
Likewise, the screening of the disclination interaction on the 
fluctuating spherical surface drives
the KT transition temperature to zero for $\mu^{-1} \ll 2$ in
which the screening length is shorter than the system size.
However, when $\mu^{-1} \gg 2$, the effect of shape fluctuations
is irrelavent and the effective KT transition occurs at the finite temperature.


\begin{thebibliography}{99}

\bibitem{jp-lub} J.M. Park and T.C. Lubensky, Phys. Rev. E {\bf 53}, 
2648 (1996); J.M. Park and T.C. Lubensky, Phys. Rev. E {\bf 53}, 
2665 (1996).

\bibitem{Nelson79} D.R. Nelson and B.I. Halperin, Phys. Rev. B {\bf 19}, 
2457 (1979).

\bibitem{Kosterlitz72} J.M. Kosterlitz and D.J. Thouless, J. Phys. C {\bf 5},
1124 (1972). J.M. Kosterlitz and D.J. Thouless, J. Phys. C {\bf 6}, 
1181 (1973). 

\bibitem{Nelson83} D.R. Nelson, ``Defect-mediated Phase Transitions'',
in {\it Phase Transitions and Critical Phenomena, Vol. 7} edited by
C. Domb and J. Lebowitz (Academic Press, New York, 1983).

\bibitem{Minnhagen87} P. Minnhagen, Rev. Mod. Phys. {\bf 59}, 1001 (1987).

\bibitem{Nelson87} D.R. Nelson and L. Peliti, J. Phys. (Paris) {\bf 48},
1085 (1987).

\bibitem{David87} F. David, E. Guitter and L. Peliti, J. Phys. (Paris)
{\bf 48}, 2059 (1987); E. Guitter and M. Kardar, Europhys. Lett. {\bf 13}, 
441 (1990).

\bibitem{Ovrut} B.A. Ovrut and S. Thomas, Phys. Rev. D {\bf 43}, 1314 (1991).

\bibitem{Helf-Canh} W. Helfrich, Z. Naturforsch {\bf 28C}, 693 (1973);
P. Canham, J. Theo. Bio. {\bf 26}, 61 (1970).

\bibitem{Lubensky92} T.C. Lubensky and J. Prost, J. Phys. II {\bf 2},
371 (1992).

\bibitem{Lubensky95} See for example 
P. Chaikin and T.C. Lubensky, {\it Principles of
Condensed Matter Physics}, (Cambridge, 1995).

\bibitem{jp-lub3} J.M. Park and T.C. Lubensky, J. Phys. I {\bf 6},
493 (1996).

\bibitem{Nelson96} M.W. Deem and D.R. Nelson, Phys. Rev. E {\bf 53},
2551 (1996).

\bibitem{Spivak79} M. Spivak, {\it A Comprehensive Introduction of
Differential Geometry}, (Publish or Perish, 1979).

\bibitem{Polyakov81} A. Polyakov, Phys. Lett. {\bf 103B}, 207 (1981).

\bibitem{Green87} M. Green, J. Schwarz and E. Witten, {\it Superstring
Theory}, (Cambridge University Press, Cambridge, 1987).

\bibitem{Moore89} M. Moore, Phys. Rev. B {\bf 39}, 136 (1989);
B.I. Halperin and D.R. Nelson, J. Low Temp. Phys.
{\bf 36}, 599 (1979).

\end{thebibliography}
\end{document}